\newif\ifjcp
\jcptrue

\ifjcp
    \documentclass[aip,jcp,reprint,floatfix]{revtex4-1}
\else
    \documentclass[journal=jacsat,manuscript=article]{achemso}
    \setkeys{acs}{articletitle = true}
    \newcommand{\onlinecite}[1]{\hspace{-1 ex} \nocite{#1}\citenum{#1}} 
\fi

\newcommand{\toadd}[1]{{#1}}
\newcommand{\toremove}[1]{\ignorespaces}

\usepackage[utf8]{inputenc}

\usepackage{natbib}
\usepackage{graphicx}
\usepackage{float}
\usepackage{physics}
\usepackage{amsmath}
\usepackage{xcolor}
\usepackage{subcaption}
\usepackage{nameref}

\newcommand{\setinfo}{%
    \title{Efficient excitations and spectra within a perturbative renormalization approach}
    \author{Oliver J. Backhouse}%
    \affiliation{Department of Physics, King's College London, Strand, London WC2R 2LS, U.K.}%
    \author{George H. Booth}%
    \email{george.booth@kcl.ac.uk}%
    \affiliation{Department of Physics, King's College London, Strand, London WC2R 2LS, U.K.}%
}

\ifjcp
    \renewenvironment{figure}{%
        \begin{figure*}%
    }{%
        \end{figure*}%
        \ignorespacesafterend%
    }
\else
\fi

\ifjcp\else
    \setinfo
\fi

\begin{document}

\ifjcp
    \setinfo
\fi

\begin{abstract}
    We present a self-consistent approach for computing the correlated quasiparticle spectrum of charged excitations in iterative $\mathcal{O}[N^5]$ computational time. 
    This is based on the auxiliary second-order Green's function approach [O. Backhouse \textit{et al.}, JCTC (2020)], in which a self-consistent effective Hamiltonian is constructed by systematically renormalizing the dynamical effects of the self-energy at second-order perturbation theory. 
    From extensive benchmarking across the W4-11 molecular test set, we show that the iterative renormalization and truncation of the effective dynamical resolution arising from the $2h1p$ and $1h2p$ spaces can substantially improve the quality of the resulting ionization potential and electron affinity predictions compared to benchmark values.
    The resulting method is shown to be superior in accuracy to similarly scaling quantum chemical methods for charged excitations in EOM-CC2 and ADC(2), across this test set, while the self-consistency also removes the dependence on the underlying mean-field reference. 
    The approach also allows for single-shot computation of the entire quasiparticle spectrum, which is applied to the benzoquinone molecule and demonstrates the reduction in the single-particle gap due to the correlated physics, and gives direct access to the localization of the Dyson orbitals.
\end{abstract}

\ifjcp
    \maketitle
\fi


The quasiparticle spectrum of a system is a key quantity in understanding its electronic structure, including its optical, magnetic, and transport properties. 
It is directly related to the one-particle Green's function on the real-frequency axis, defining how we probe electronic systems via the addition or removal of an electron and subsequent observation of the response of the system to this change. 
Experimentally, this is performed via direct/inverse photoelectron spectroscopy or scanning tunnelling microscopy, and computational tools to predict and simulate this process are of great importance in many fields where charge transfer processes are probed. 
This includes areas ranging from organic to atmospheric chemistry \cite{Mohen1976}, development of photovoltaics and dyes \cite{Eriksson2016,Snook2005}, as well as determining the band structure edges in extended systems \cite{Barr1989,Iskakov2019}.

This spectrum exhibits poles corresponding to ionization and electron attachment events in such an experiment, each with an associated transition amplitude, and can be computed according to
\begin{equation}
    A(\omega) = \frac{1}{\pi} \mathrm{Im} [ G(\omega + i\eta) ], \label{eq:QPSpectrum}
\end{equation}
where $G(\omega)$ is the frequency-dependent one-particle Green's function of the system, and $\eta$ is a formally infinitesimally small positive number.
The determination of accurate molecular ionization potentials (IPs) and electron affinities (EAs) can in principle be calculated at any level of theory by performing separate ground-state calculations on the $(N+1)$- and $(N-1)$-electron reference states, and observing the difference in energies with respect to the $N$ electron system. Such methods are often denoted using a $\Delta$ prefix combined with the level of theory, such as $\Delta$MP2, which shall be used in the present work.
These methods can perform well as there is a necessary cancellation of many errors in the given level of theory, however they are limited to only computing the lowest energy IP/EA and are not capable of providing transition moments in a simple way, both of which are crucial for determining the quasiparticle weight and comparison to spectroscopic experiments. 
Alternatively, these ionized/electron-attached states can be considered as charged excited states of the system, and methods have been developed to directly target these excitation energies. 
These are important tools in quantum chemistry, and include the equation-of-motion coupled cluster (EOM-CC)\cite{Monkhorst1977,Stanton1993,Stanton1994} and methods derived via Green's function methods including the $GW$ method \cite{Hedin1965,Hybertsen1985,Aryasetiawan1998,vanSetten2013,Kaplan2016}, algebraic diagrammatic construction \cite{Schirmer1983,Trofimov1995,Trofimov1999} (ADC), and second-order Green's function perturbation theory \cite{Holleboom1990,VanNeck2001,Dahlen2005,Phillips2014} (GF2).

In this work, we explore the accuracy and suitability of the `auxiliary second-order Green's function perturbation' approach (AGF2) for predicting quasiparticle spectra, recently developed by the authors in Ref.~\onlinecite{Backhouse2020}. 
This is a fully self-consistent method, which iteratively renormalizes the Dyson orbitals and single-particle spectrum of the system in response to the correlations at second-order perturbation theory. 
In this way, the method is identical to the aims of traditional GF2, however, the AGF2 approach operates in a grid-free formalism, which enables \toadd{individual IPs/EAs to be found without numerical ambiguities from analytic continuation, and allowing for direct} comparison to excited state quantum chemical methods. 
\toremove{More importantly, AGF2 also has a combination of two additional physically-motivated approximations, related to the systematic renormalization and truncation of implicit dynamical effects.}
\toadd{The method employs a frequency-independent auxiliary space, whose construction and iterative optimization defines an reduced dimensionality effective hamiltonian for the charged excitations each iteration. Initially, these auxiliaries are just constructed from the $2h1p$ and $1h2p$ spaces of a reference (single determinant) state, however through the renormalization of these auxiliary states and subsequent self-consistency, this excitation character of the auxiliary states is lost, as they now correspond to excitations from an enlarged ficticious system. Instead, these auxiliaries can be related to the dynamical character of the self-energy in GF2, by implicitly mimicking the frequency-dependence of the iterative self-energy.
Importantly, AGF2 also has a combination of two physically-motivated approximations, related to how this systematic renormalization and truncation of the auxiliaries is defined, and therefore of the implicit dynamical self-energy effects.}
We will explore the effect of these truncations, and compare and contrast the approach to alternate methods in quantum chemistry for computing charged excitations. 
This is done with extensive benchmarking to the W4-11 molecular test set, where we find AGF2 with this dynamical renormalization provides highly competitive charged excitation energies, outperforming other $\mathcal{O}[N^5]$ approaches.

\section{Charged excitations in quantum chemistry}

In this section, we briefly describe the salient features of the more common approaches within quantum chemistry for the calculation of charged excitations. 
This allows for an understanding of the differences to the AGF2 approach of this work, and includes references for further information.

Equation of motion coupled cluster (EOM-CC) \cite{Monkhorst1977,Stanton1993}, expanded to permit computation of non-neutral excitations by Stanton and Gauss \cite{Stanton1994}, is one of the most commonly applied quantum chemical methods in the determination of IPs and EAs, along with the highly related linear-response coupled cluster\cite{Dalgaard1983,Koch1990}.
As with the parent method, this is most commonly performed to EOM-CCSD level and scales with $\mathcal{O}[N^6]$, as it first requires the calculation of the ground state (CCSD) solution. 
There also exist a number of approximations to include partial information of triples contributions at the cost of higher scaling \cite{Dutta2017}. 
One such approximation by Matthews and Stanton closely resembles the asymmetric triples approximation to the CCSD(T) correction in the ground-state energy, with a further approximation in the ionized (or excited) state \cite{Matthews2016}.
This correction to EOM-CCSD is termed EOM-CCSD(T)(a)* and approaches the quality of full EOM-CCSDT at a reduced cost of $\mathcal{O}[N^7]$, and serves for many of the benchmark comparison values in the present work.

Equally, there also exists a series of more approximate lower-scaling alternatives to the EOM-CCSD method, one of the most popular being the approximate singles and doubles method EOM-CC2.
CC2 approximates the doubles equation of CCSD to only be correct through first-order, with the singles as zeroth-order parameters, reducing the overall scaling down to $\mathcal{O}[N^5]$, and avoiding explicit renormalization of fixed MP2-like $\hat{T}_2$ amplitudes.
The performance of CC2 for vertical excitation energies is considered to be good, given the modest scaling, and has been extended to large molecular systems making use of the resolution of identity (RI) approximation \cite{Hattig2000,Sneskov2012}.

As an alternative, the algebraic diagrammatic construction (ADC) method, traditionally motivated via a diagrammatic expansion of the polarization propagator or Green's function, is an excited-state counterpart to M{\o}ller-Plesset perturbation theory \cite{Schirmer1983,Trofimov1999}. 
The second-order ADC(2) (scaling as $\mathcal{O}[N^5]$), extended second-order ADC(2)-x (formally $\mathcal{O}[N^6]$, $\mathcal{O}[N^5]$ without transition moments\cite{Banerjee2019}), and third-order ADC(3) ($\mathcal{O}[N^6]$) methods have seen some recent popularity in the literature for calculating ionization potentials and electron affinities, as well as excitation energies \cite{Mester2018,Banerjee2019,Hodecker2020,Herbst2020}.
Using an iterative eigensolver ADC($n$) is capable of calculating these quantities, including transition amplitudes, at a level consistent through order-$n$ perturbation theory \cite{Weikert1996}.
Results have been shown to be good given the computational scaling of the methods, with the $\mathcal{O}[N^6]$ scaling ADC(3) method performing at a similar level to the equally scaling EOM-CCSD method (however ADC(3) can be made generally more efficient due to its non-iterative nature) \cite{Banerjee2019}.
In ADC(2), the $1h2p$/$2h1p$ block is diagonal since it is expanded only to zeroth order in the perturbation, resulting in the lower scaling of the approach, at $\mathcal{O}[N^5]$.
Just as increasing the order of expansion of ADC(2) defines ADC(3), if one only increases the order of the $1h2p$/$2h1p$ block, the ADC(2)-x method is defined.
Consequently, this flavour of ADC does not have rigorous theoretical justification in perturbation theory, and provides an unbalanced descriptor which consistently underestimates excitation energies \cite{Harbach2014}. 
The effective Hamiltonian of ADC(2) is also closely related to the Jacobian of CC2, via symmetrization of this Jacobian and removal of the $\hat{T}_1$ amplitudes \cite{Dreuw2015,Christiansen1995}, justifying the similar level of accuracy between these methods.

Whilst ADC has shown some success in the calculation of excited states, it is not without pitfalls.
Owing to its role as an excited-state version of M{\o}ller-Plesset perturbation theory, it shares the limitations of the underlying MP$n$ ground state (compared to the equivalent CC$n$ ground state), including the poor description at points of stronger correlation effects \cite{Banerjee2019}.
Successful ADC calculations therefore require a sensible description of this ground state, and cannot be expected to perform well if this prerequisite is not met \cite{Dreuw2015}. 
Both ADC and EOM-CCSD for charged excitations also exhibit a strict separation between hole (IP) and particle (EA) excitations, with these uncoupled and able to be solved entirely independently. 
Once separated, both ADC and EOM-CCSD approaches build effective Hamiltonians for the perturbed states, and iteratively solve for the extremal eigenvalues and vectors corresponding to the lowest energy IPs and EAs. 
While these are often most relevant in chemical physics, this approach precludes a state-specific approach for construction of the entire quasiparticle spectrum, or even straightforward access to important deep-lying core excitations in the interior of the effective Hamiltonian, as complete diagonalization would scale prohibitively \cite{Barth1981,Barth1985,Wenzel2015}. 
If the full spectrum is required, then GF-CCSD approaches build the response, targeting and solving for a single frequency at a time \cite{Nooijen1995,Shee2019,Zhu2019}, but this removes the description of individual, state-specific excitations.

In contrast, Green's function methods such as GF2 or $GW$ solve for the entire spectrum, often without a clearly defined ground state wave function.
\toremove{Optional self-consistency} \toadd{Self-consistency} on the level of the underlying Green's function can change the zeroth-level description due to the entire set of charged excitations, via the introduction of a self-energy \toadd{and associated correlation effects}.
\toadd{These changes to the propagator are computed via the Dyson equation, which connects the Green's function and self-energy.}
\toadd{Whilst one-shot corrections such as $G_0W_0$ are common, they are not always rigorously justified from a functional perspective and lack appropriate conservation laws.}
\toremove{These} \toadd{Green's function} schemes \toremove{however} \toadd{also} generally require the introduction of a time or frequency variable, which for numerical reasons is not generally taken to be on the real-frequency axis where it is required for the physical spectrum in Eq.~\ref{eq:QPSpectrum}.
This necessitates the use of analytic continuation techniques \toremove{if the bandstructure is fully renormalized,} which removes the accurate specification of particular IP or EA excitations \cite{Welden2015}.

\section{Auxiliary Green's function second-order perturbation theory} \label{sec:AGF2}

Auxiliary Green's function second-order perturbation theory (AGF2) was presented in Ref.~\onlinecite{Backhouse2020}, where the algorithm was described and benchmarking of ground-state energies was performed. 
The approach however naturally lends itself to the computation of the charged excitations, which will be considered here in more detail. 
Readers interested in the methodology and formulation may refer to our previous paper, here we will only recap the main points in its construction which are particularly relevant to the present work.

The method brings together the Green's function and `effective Hamiltonian' approaches of ADC and EOM-CC described above. 
From the perspective of an effective Hamiltonian model, the method constructs an effective Hamiltonian with a similar structure to ADC(2). 
However, the effect of the $2h1p$ and $1h2p$ spaces is first systematically compressed into a set of contracted and renormalized `auxiliary' degrees of freedom, whose dimension is rigorously $\mathcal{O}[N]$. 
This space is then combined with the explicit $1h$ and $1p$ spaces, and completely diagonalized, made possible by its compact size. 
The renormalization of the $1h$ and $1p$ spaces due to these auxiliary states is then used to construct a new $2h1p$ and $1h2p$ space, which is again compressed, and the whole procedure iterated to convergence. 
In this way, at each stage the entire spectrum of charged excitations is found, giving access to the entire quasiparticle spectrum. 
Furthermore, separation of particle and hole excitations is not performed in the self-consistency, and full renormalization of the underlying (Dyson) orbitals and bandstructure in response to the correlations over all energy scales is performed, in contrast to effective Hamiltonian approaches where the ground-state description is not updated after the excitations are found. 
In \toremove{Sec.~\ref{sec:results}} \toadd{the results} we discuss further the nature and importance of this self-consistency in the description of the excitation energies.

The AGF2 approach can also be motivated from a renormalized Green's function perspective, where it can be considered to be a systematic renormalization and truncation of the implicit dynamical resolution of the GF2 method on the real-frequency axis \cite{Backhouse2020}.
\toadd{This `dynamical resolution' corresponds to the degree of coarse-graining of the $2h1p$ and $1h2p$ spaces, which is renormalized to just ensure that it represents a desired number of moments of the Green's function and self-energy.}
The method begins with the frequency-domain 2nd-order perturbative self-energy in the MP2 partitioning, given as
\ifjcp
    \begin{align}
        \label{eq:mp2_self_energy}
        \Sigma_{pq}(\omega) &= 
        \sum_{ija} \frac{(pi|ja)[2(qi|ja)-(qj|ia)]}{\omega-E_{i}-E_{j}+E_{a}} 
        \\ &+ 
        \nonumber
        \sum_{abi} \frac{(pa|bi)[2(qa|bi)-(qb|ai)]}{\omega-E_{a}-E_{b}+E_{i}},
    \end{align}
\else
    \begin{equation}
        \label{eq:mp2_self_energy}
        \Sigma_{pq}(\omega) = 
        \sum_{ija} \frac{(pi|ja)[2(qi|ja)-(qj|ia)]}{\omega-E_{i}-E_{j}+E_{a}} +
        \sum_{abi} \frac{(pa|bi)[2(qa|bi)-(qb|ai)]}{\omega-E_{a}-E_{b}+E_{i}},
    \end{equation}
\fi
where $E$ are the Hartree--Fock eigenvalues, and $(ij|kl)$ are two-electron integrals in chemists' notation.
One then solves the Dyson equation $G = G_0 + G_0 \Sigma G$ which represents a dressing of the Green's function $G$, according to the irreducible correlations defining $\Sigma$. 
In our notation, $p,q$ denote physical degrees of freedom, $i,j$ are occupied molecular orbitals, and $a,b$ virtual molecular orbitals (MOs).
However in AGF2, rather than performing the Dyson equation explicitly, we define the action of this self-energy as an effective Hamiltonian coupling the $1h/1p$ Hamiltonian sector (Fock matrix) to a $2h1p/1h2p$ sector.
These terms are equivalent to those defined in ADC(2) (neglecting first-order terms, which are iteratively rather than perturbatively included in this approach).
In ADC(2)/EOM-CCSD however, the $1h$ couples only to the $2h1p$ sector and the $1p$ to the $1h2p$, whereas in AGF2, additional couplings are included between all spaces (including e.g. coupling between the $1h$ and $1h2p$ spaces), since the $p,q$ indices of Eq.~\ref{eq:mp2_self_energy} run over all physical orbitals.  

The states generated in these $2h1p$ and $1h2p$ sectors are considered as a set of separable auxiliary states coupling to the `physical orbitals', whose Hamiltonian is the one-electron Fock matrix of the system. These auxiliary states can be considered as implicitly representing an effective self-energy given by
\begin{equation}
    \label{eq:aux_self_energy}
    \Sigma_{pq}(\omega) = 
    \sum_{\alpha} \frac{v_{p\alpha} v_{q\alpha}^\dagger}{\omega-\epsilon_\alpha},
\end{equation}
where the auxiliaries have energies $\epsilon$ and couple to the physical system according to couplings $v$.
The Dyson equation can then be reformulated as an eigenvalue problem of an {\em extended Fock matrix} as
\begin{equation}
    \label{eq:dyson_eigenvalue}
    \begin{bmatrix} & F & v & \\ & v^\dagger & \mathrm{diag}(\epsilon) & \end{bmatrix} \phi = \lambda \phi,
\end{equation}
where $F$ is the Fock matrix, and $\phi$ we called quasi-molecular orbitals (QMOs) with energies $\lambda$, which span the physical and auxiliary system, allowing the weight on each orbital in the physical space to change from its Hartree--Fock value.
These QMOs define Dyson orbitals once projected into the physical space \cite{Oana2007}, and can then be used to define a new self-energy by replacing the physical MOs in Eq.~\ref{eq:mp2_self_energy} by the QMOs which span both the physical and auxiliary degrees of freedom, defining a self-consistent procedure, which is equivalent to a real-frequency self-consistent GF2 calculation \cite{VanNeck2001}.

Unchecked, the number of these auxiliaries in this procedure grows cubically with each iteration, and therefore a truncation in this auxiliary space is required such that it does not increase in dimensionality through the iterations and remains computationally tractable.
This is achieved via a renormalization of the auxiliary space, which is defined such that the resulting truncated auxiliary space represents the effect of a self-energy and subsequent Green's function with an identical set of spectral moments in their frequency distribution as the original $\Sigma(\omega)$ and subsequent $G(\omega)$.
This renormalization ensures that the number of auxiliary states scales linearly with the number of physical orbitals of the system by construction. It is worth clarifying the meaning of the term `renormalization', which is used in two contexts in this work. The Green's function is iteratively renormalized in a sense that the propagator is dressed via a self-energy, as in traditional GF2. However, this self-energy is also subject to a renormalization step each iteration, through the contraction and truncation of the auxiliary space, in the manner of the density matrix renormalization group \cite{White1992,Schollwock2011}, as described above.

AGF2 calculations are therefore denoted AGF2($n_\mathrm{mom}^G$,$n_\mathrm{mom}^\Sigma$), where $n_\mathrm{mom}$ are the orders of the matching in the respective spectral moments of each quantity.
Renormalization of the auxiliaries to order $n_\mathrm{mom}^\Sigma$ ensure consistency in the self-energy moments before and after truncation up to order $2n_\mathrm{mom}^\Sigma+1$ in the separate occupied and virtual sectors of the self-energies, and similar constraints apply for $n_\mathrm{mom}^G$ in the truncation of the effective dynamical resolution of the resulting Green's function of the system. Other metrics for defining the renormalization of this auxiliary space can also be formulated \cite{Nusspickel2020}.

The $n_\mathrm{mom}^\Sigma$ truncation was first performed for self-consistent Green's function methods in the nuclear physics community where it was termed `BAGEL' \cite{Muther1988,Muther1993,Dewulf1997}, and applied previously within a self-consistent GF2 approach by Van Neck \textit{et al.} \cite{VanNeck2001,Piers2002}. 
We note that the AGF2(1,0) approach, which is shown to be an accurate truncation in the present work, is very similar to the BAGEL(1,1) approach, with only differences in the method of solution of the Dyson equation. 
However, in the original work this truncation was not investigated for spectral properties, rather ground-state energies, for which we showed that higher moment truncations are superior \cite{Backhouse2020}. 
Similarly, truncations in the moments of the Green's function have been investigated recently in strong coupling embedding approaches as a method for systematically truncating the effect of a hybridization \cite{Fertitta2018,Fertitta2019}.

Once these two truncations in $n_\mathrm{mom}^\Sigma$ and $n_\mathrm{mom}^G$ are chosen (which approximates the effective dynamical resolution of the quantities via renormalization of the auxiliary space), the scaling of the method with system size is only $\mathcal{O}[N^5]$, with higher truncations only increasing the prefactor to these calculations.
There is also an additional renormalization of the propagator in the one-body diagrams each iteration, via an update of the Fock matrix as the (non-idempotent) density relaxes due to the presence of the updated auxiliary space.
\toadd{As such, the resulting Fock matrix is non-diagonal, in contrast to that of the original mean-field calculation.
Whilst this step has an appreciable prefactor, it scales with system size as $\mathcal{O}[N^4]$ and therefore is not a computationally rate-limiting step.
Convergence of this self-consistent loop is rarely problematic, and DIIS is employed to accelerate this convergence.
The specific steps which contribute to the method performing with an $\mathcal{O}[N^5]$ scaling are the three-quarter transformation of the electronic repulsion integrals into the QMO basis, and the $n_\mathrm{mom}^\Sigma$ truncation.
The prefactor of the former depends on $n_\mathrm{mom}^G$ and the latter on $n_\mathrm{mom}^\Sigma$, with both operations required at every iteration of AGF2.
This is likely to be an increase on the number of operations required by the iterative diagonalisation in ADC(2), however AGF2 calculations produce a representation of the entire spectrum of excitations, rather than just the lowest-lying states.
AGF2 also allows for the mixing of the $2h1p$ and $1h2p$ states rather than allowing their strict separation as in CC2 and ADC(2).}


The frequency-dependent Green's function can be built using the converged QMOs according to
\begin{equation}
    \label{eq:aux_greens_function}
    G_{pq}(\omega) = \sum_{x} \frac{\phi_{px} \phi_{qx}^\dagger}{\omega-\lambda_{x}},
\end{equation}
where $x$ labels the entire set of QMOs.
This provides direct access to the real-frequency spectral function $A(\omega)$ without the need for numerical grids.
In addition, the exact energies of all charged excitations, their transition amplitudes and Dyson orbitals are all easily accessible via $\lambda$ and the component of $\phi$ projected in the physical space.

Unlike CC2 and ADC(2), GF2 (and thus AGF2) does not {\em a priori} require a qualitatively correct description of the mean-field ground state in order to be accurate, despite still being motivated via the MP2 method \cite{Phillips2014}.
This owes to the renormalization of the \toadd{first- and} second-order diagrams through the iterations, which does not occur in ADC(2) or CC2, the former being non-iterative and the latter only renormalizing $\hat{T}_1$ \cite{Christiansen1995}. 
This ensures that GF2/AGF2 is formally independent of the mean-field reference state it starts from.
We therefore hope that the renormalization of these diagrams provided by AGF2 can benefit the calculation of IPs and EAs, just as it benefits the ground state energy.

In more traditional GF2 approaches, one typically performs the calculation on the Matsubara and imaginary time domain, and therefore must perform an ill-posed and numerically unstable inverse Laplace transform to the real frequency axis to analytically continue the spectral function \cite{Kraberger2017}.
This makes it very difficult to access accurate excitation energies, despite formally having all the information about these quantities.
In an attempt to circumvent this analytic continuation, Welden {\em et. al.} investigated the use of the extended Koopmans' theorem (EKT) within GF2 in order to extract the IPs and EAs of the system\cite{Welden2015}. 
This method approximates the IP/EA via a diagonalization of an effective Hamiltonian generated by the first moment of the Green's function. 
This is similar in spirit to the Green's function moment truncations performed in this work, however in AGF2 the Green's function is truncated at each iteration to this effective dynamical resolution via renormalization of the auxiliary space rather than a fully self-consistent dynamical Green's function being truncated in order to compute the excitation energies. 
Nevertheless, Welden \textit{et al.} show that the IPs and EAs of fully dynamical GF2 are systematically underpredicted, and that limiting the GF2 self-consistency (by taking only the first iteration Green's function) improves them, as also noted by Dahlen and van Leeuwen prior to this work \cite{Dahlen2005,Stan2009}. 
The present work will build upon these conclusions, making use of the systematic and self-consistent effective dynamical truncations inherent in AGF2 and the ability to compute the IPs and EAs without approximating them via the EKT, to develop an efficient and accurate procedure for their calculation within this scheme.
This would allow access to a method which retains the $\mathcal{O}[N^5]$ scaling whilst computing spectral properties at a fully renormalized perturbative level more closely resembling that of higher-order methods, such as EOM-CCSD and EOM-CCSD(T).


\section{Results and discussion}\label{sec:results}

We will benchmark results using the W4-11 test set, consisting of a series of 140 neutral molecules, including some with small levels of static correlation effects, such as C$_2$ \cite{Karton2011}. 
To avoid complications of symmetry breaking and choice of reference, we restrict the set to only include the closed-shell species, and will return to consider open-shell AGF2 performance in later work. 
\toadd{Additionally, to permit use of more expensive levels of theory, we limit the test set to molecules with fewer than 50 basis functions in a cc-pVDZ basis, leaving 68 molecules in the set.}
The calculations were performed in a cc-pVDZ basis set, with all electrons correlated.
IP/EA-EOM-CC2 calculations were performed using the CFOUR package \cite{cfour,Christiansen1995}, whilst all other calculations except AGF2 were calculated using PySCF \cite{pyscf,Banerjee2019}.
The AGF2 calculations were performed using an in-house program building on existing integral and mean-field functionality of PySCF.

\subsection{Effect of auxiliary space renormalization in self-consistent AGF2}

We first consider the effect of the effective dynamical truncations via renormalization of the auxiliary space, both in the effective system Green's function ($n_\mathrm{mom}^G$) and self-energy ($n_\mathrm{mom}^\Sigma$), on the accuracy of the IP and EA values from self-consistent AGF2. 
The level of this truncation determines the resulting number of auxiliary states in the effective Hamiltonian of Eq.~\ref{eq:dyson_eigenvalue} at each iteration, with lower truncation levels leading to a more aggressive contraction of these auxiliary states. 
For comparison, we also include results from a non-self-consistent AGF2 calculation, where only a single self-energy is computed, giving the spectral values due to a single-shot MP2 self-energy. 
This results from a single iteration of the algorithm, and will be termed AG$_1$F2 in keeping with prior literature \cite{Welden2015}. This will also allow us to determine the importance of self-consistency alongside the effective dynamical resolution of the key quantities. 
Unlike any of the self-consistent Green's functions, AG$_1$F2 will remain reference-dependent, with diagrammatic contributions which are not iteratively renormalized at this level.
However, while the two-body effects are not included in the dressing of $G(\omega)$, the density is fully relaxed due to the presence of the two-body correlations via a self-consistent optimization of one-body contributions in the Fock matrix defining the propagator. 
This ensures that AG$_1$F2 corresponds to a single iteration \toremove{of the procedure outlined in Sec.~\ref{sec:AGF2} and our previous work} \toadd{of the method, as outlined in our previous work} \cite{Backhouse2020}. This full relaxation of the one-body diagrams, along with the additional self-energy couplings between the $1h$-$1h2p$ and $1p$-$2h1p$ sectors are the only formal differences between this AG$_1$F2 approach and ADC(2), and they are shown later to have similar accuracy \toadd{when high truncation levels are used in the AGF2}.

\begin{table}
    \centering
    \begin{tabular}{c c c}
        \hline
       Truncation & MAE (IP) & MAE (EA) \\
        \hline                            
   AG$_{1}$F2 &        0.832 &        0.479 \\
    AGF2(0,0) &        0.325 &        1.175 \\
    AGF2(0,1) &        0.286 &        1.231 \\
    AGF2(0,6) &        0.502 &        1.268 \\
    AGF2(1,0) &        0.299 &        0.217 \\
    AGF2(1,1) &        0.546 &        0.275 \\
    AGF2(1,6) &        1.031 &        0.382 \\
    AGF2(2,1) &        0.524 &        0.424 \\
    AGF2(5,6) &        1.303 &        0.926 \\
        \hline
    \end{tabular}
    \caption{Mean absolute errors (MAEs) compared to EOM-CCSD(T)(a)* for the IPs and EAs of the W4-11 test set (eV). The nomenclature for effective dynamical truncation of self-consistent AGF2 is $(n_\mathrm{mom}^G$,$n_\mathrm{mom}^\Sigma)$.}
    \label{tab:truncation_errors}
\end{table}

\begin{figure}
    \centering
    \includegraphics[width=\textwidth]{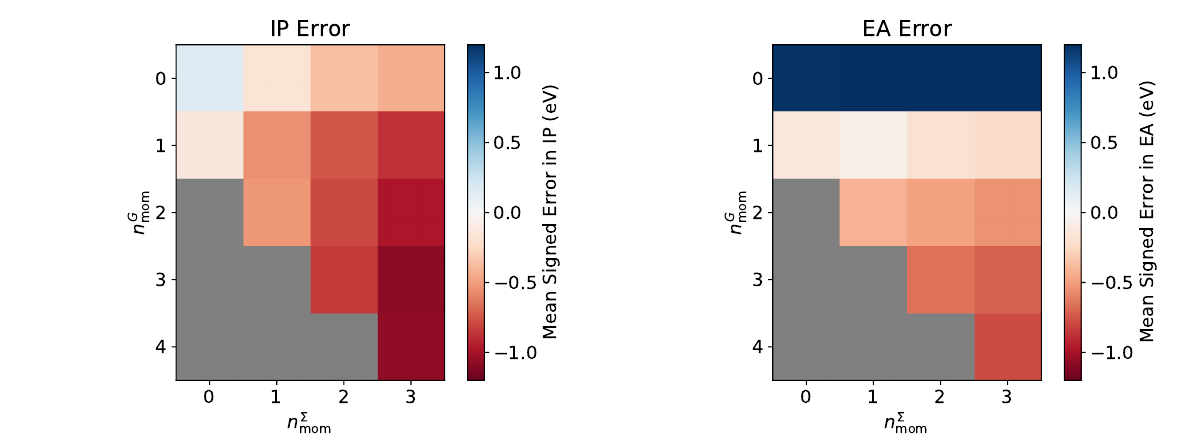}
    \caption{Heat maps showing the mean (signed) error in the IP (left) and EA (right) for different effective dynamical truncations of self-consistent AGF2. This is benchmarked against EOM-CCSD(T)(a)* for the W4-11 set.}
    \label{fig:w4_contour}
\end{figure}

In Table~\ref{tab:truncation_errors} and Fig.~\ref{fig:w4_contour}, we show the errors for a range of systematic truncations of the self-consistent effective dynamics of the Green's function and self-energy in AGF2, as well as comparison in Table~\ref{tab:truncation_errors} to non-self-consistent AG$_1$F2. 
These errors are taken with respect to benchmark EOM-CCSD(T)(a)* values. 
Mean absolute errors in Table~\ref{tab:truncation_errors} ensure that fortunate error cancellation in aggregated data sets is avoided, while the signed errors of Fig.~\ref{fig:w4_contour} show the systematic over- or under-estimation of these excitation energies. 
The ``$(5,6)$'' truncation of AGF2 is essentially converged with respect to the effective dynamical quantities, and can be considered to be the real-frequency self-consistent GF2 limit.

As also concluded in other studies and discussed \toremove{in Sec.~\ref{sec:AGF2}} \toadd{previously}, for both the IP and EA, the performance of AGF2(5,6) in this `fully dynamical' self-consistent limit is generally very poor and even worse than Hartree--Fock, with a significant underestimation in both the IP and EA across the W4-11 test set\cite{Welden2015}. 
In fact, for many of the larger moment truncations, the non-self-consistent AG$_1$F2 performs better than the self-consistent solution, which is in agreement with conclusions made by Welden \textit{et al.} \cite{Welden2015}. 
However, this is not necessarily true of the more computationally tractable lower-moment AGF2 calculations with coarser resolution of the effective dynamics and smaller auxiliary spaces, where self-consistency is found to significantly improve results.
Figure~\ref{fig:w4_contour} clearly shows that for accurate IPs, the importance appears to lie in keeping both $n_\mathrm{mom}^G$ and $n_\mathrm{mom}^\Sigma$ low, whilst the EA is less dependent on $n_\mathrm{mom}^\Sigma$, and requires only that $n_\mathrm{mom}^G=1$.
AGF2(0,0), AGF2(0,1) and AGF2(1,0) all perform with impressive accuracy for the IP, and AGF2(1,0) and AGF2(1,1) for the EA, while $n_\mathrm{mom}^G=0$ overestimates the EA values.

The fact that fully self-consistent approaches with the full effective dynamical information deteriorates the quality of these excitation energies is not a unique feature of AGF2, and has been observed in many self-consistent Green's function approaches. 
The inclusion of additional diagrams in models does not guarantee a more accurate method due to an unbalanced description of the neutral ground and excited state, and needs to be carefully considered. 
For instance, the $GW$ method is one of the most widespread beyond-mean-field Green's function approaches in condensed matter. 
In this method, full dynamical self-consistency has been found to reduce the quality of results for bandgap edges, and instead limited dynamical self-consistency or even no self-consistency has proven to be necessary for improved results \cite{vonBarth1996,Schone1998,vanSchilfgaarde2006,vanSetten2013}. 
Even in quantum chemistry applications, these situations also abound, with ADC(2)-x performing worse for excitation energies than the ADC(2) method (despite a formally more complete diagrammatic expansion) \cite{Harbach2014}, and the CC2 method performing well for excitation energies relative to ground state energetics \cite{Pabst2010}.

The AGF2(1,0) truncation scheme therefore appears to be optimal in terms of accuracy of both IP and EA quantities across the W4-11 database, is fully self-consistent, and is also exceptionally efficient and simple to implement. 
This approach consists of a single iteration of the block Lanczos algorithm for compression of the auxiliary space at each iteration, in order to fulfil the requirement of matching only $n_\mathrm{mom}^\Sigma=0$ (which ensures that the effective self-energy at each iteration is retained up to the first spectral moment in the hole and particle sectors). 
This renormalizes the number of auxiliary states at each iteration to only have a maximum number of twice the dimensionality of the number of orbitals in the system, and is therefore particularly efficient. 
The additional $n_\mathrm{mom}^G=1$ constraint does therefore not in practice change the auxiliaries further to this, as this constraint is less restrictive than the self-energy constraint.
Having such a simple truncation scheme, it is a very attractive candidate for a competitive $\mathcal{O}[N^5]$ method for spectral properties, with mean absolute errors (MAEs) for the IP and EA of 0.299 and 0.217 eV respectively.
We therefore progress to considering the performance of the AGF2(1,0) method for charged excitations against other approaches, to assess its accuracy in a wider context.

\subsection{Benchmarking accuracy of AGF2(1,0)}

%
%

\begin{figure}
    \centering
    \includegraphics[width=\textwidth]{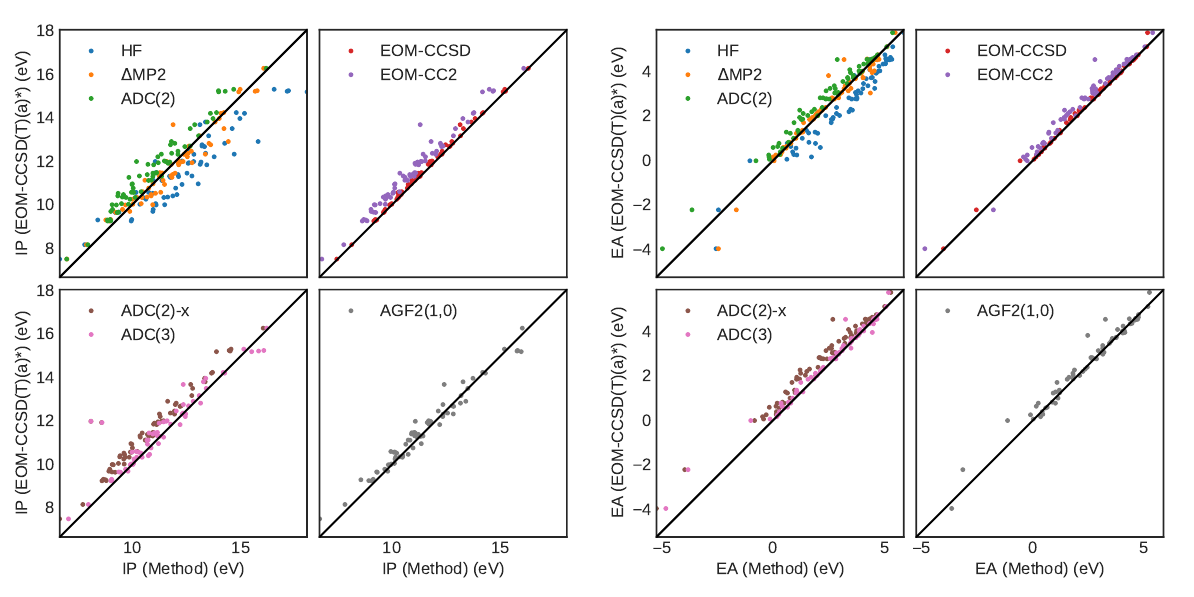}
    \caption{Correlation in the IP (left) and EA (right) between studied methods and EOM-CCSD(T)(a)* reference for the W4-11 set \toadd{in a cc-pVDZ basis}, where scatter points closer to the black lines indicate better accuracy.}
    \label{fig:w4_correlation}
\end{figure}

\begin{table}
    \centering
    \begin{tabular}{l c c}
        \hline
       Method & MAE (IP) & MAE (EA) \\
        \hline                            
           HF &       0.902 &        0.698 \\
  $\Delta$MP2 &       0.357 &        0.197 \\
      EOM-CC2 &        0.611 &        0.376 \\
     EOM-CCSD &        0.105 &        0.094 \\
       ADC(2) &        0.603 &        0.376 \\
     ADC(2)-x &        0.775 &        0.507 \\
       ADC(3) &        0.371 &        0.183 \\
    \hline
    AGF2(1,0) &        0.299 &        0.217 \\
        \hline
    \end{tabular}
    \caption{Mean absolute errors (MAEs) between each method and EOM-CCSD(T)(a)* for the IPs and EAs in a cc-pVDZ basis (eV).}
    \label{tab:w4_errors}
\end{table}

\begin{figure}
    \centering
    \includegraphics[width=\textwidth]{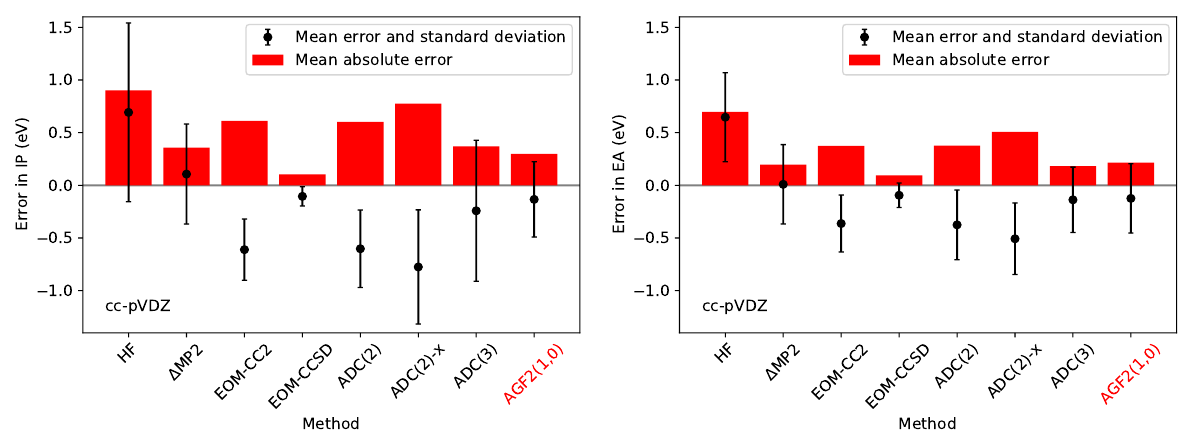}
    \caption{Summary of mean (signed) errors and their standard deviations (black error bars) for the IP (left) and EA (right) across the W4-11 test set \toadd{in a cc-pVDZ basis} compared to EOM-CCSD(T)(a)*. In addition, the mean absolute errors are shown as red bars for each studied method.}
    \label{fig:w4_summary}
\end{figure}


\begin{figure}
    \centering
    \includegraphics[width=\textwidth]{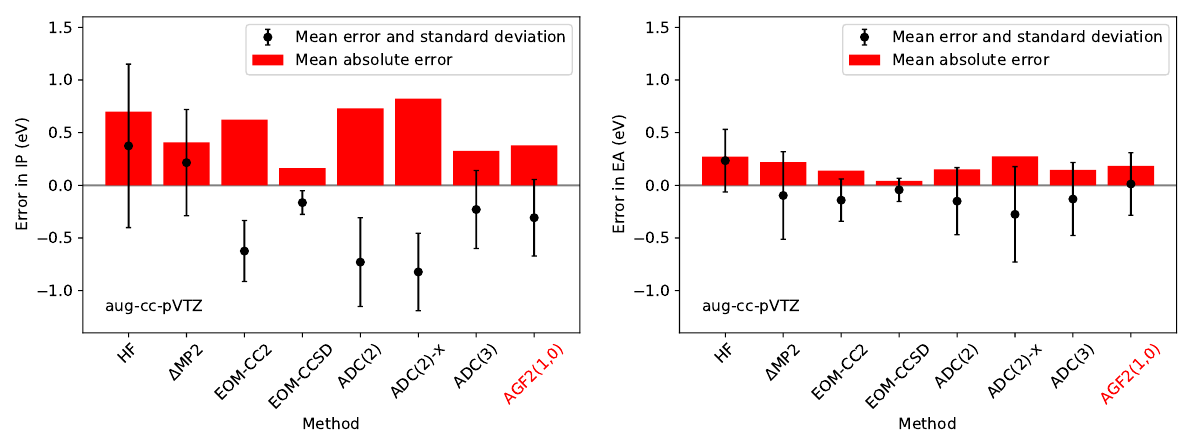}
    \caption{Summary of mean (signed) errors and their standard deviations (black error bars) for the IP (left) and EA (right) across the W4-11 test set \toadd{in an aug-cc-pVTZ basis} compared to EOM-CCSD(T)(a)*. In addition, the mean absolute errors are shown as red bars for each studied method.}
    \label{fig:w4_summary_augccpvtz}
\end{figure}

Fig.~\ref{fig:w4_correlation} shows a correlation plot giving the error for the IP and EA of each comparison method to the benchmark EOM-CCSD(T)(a)*, for each system studied in the W4-11 database. 
The further the scatter is from the diagonal, the larger the error that is made with respect to this reference. 
These results are aggregated into mean absolute errors for each method in Table~\ref{tab:w4_errors}, while both the mean and standard deviation of the (signed) errors, as well as the mean absolute error are summarized graphically in Fig.~\ref{fig:w4_summary}.
IPs and EAs calculated at the Hartree--Fock level make use of Koopmans' theorem and neglect orbital relaxation, despite the LUMO often being a poor approximation to the EA.
The accuracy of Koopmans' theorem in estimating EAs is well documented in the literature, though is likely to be aided in this study by the somewhat restricted basis set\cite{Heinrich1986}.

Predictably, the $\mathcal{O}[N^6]$ scaling methods which are consistent through higher orders of perturbation theory such as EOM-CCSD and ADC(3) perform the best. Poor performance is observed for Hartree--Fock (HF), which tends to overestimate the value of both the IP and EA, leading to a substantial overestimation of the fundamental gap, as is well known in the literature.
ADC(2), ADC(2)-x and EOM-CC2 generally underestimate both of these quantities, whilst $\Delta$MP2 performs reasonably well, however does not allow the calculation of transition moments and/or higher-lying excited states or spectral quantities.
The performance of these methods on this test set is in good quantitative agreement with a study by Walz \textit{et al.}\cite{Walz2016} on the hierarchy of coupled cluster singles and doubles methods for IPs.
With the exception of the EA for $\Delta$MP2 for which the accuracy is similar, AGF2(1,0) outperforms all other $\mathcal{O}[N^5]$ methods included in this study over this test set ($\Delta$MP2, EOM-CC2, ADC(2)). 
AGF2(1,0) in fact achieves essentially the same level of accuracy as ADC(3), a (non-iterative) $\mathcal{O}[N^6]$ scaling method, for which the MAEs in IP and EA are 0.371 and 0.183 eV respectively.
The ADC(3) method is complete through a higher level of perturbation theory (third-order), and so it is encouraging to see AGF2(1,0) reach similar levels of error despite the aggressive renormalization of the auxiliary space and therefore coarse-graining of the effective dynamics in this self-consistent method.

\begin{figure}
    \centering
    \includegraphics[width=\textwidth]{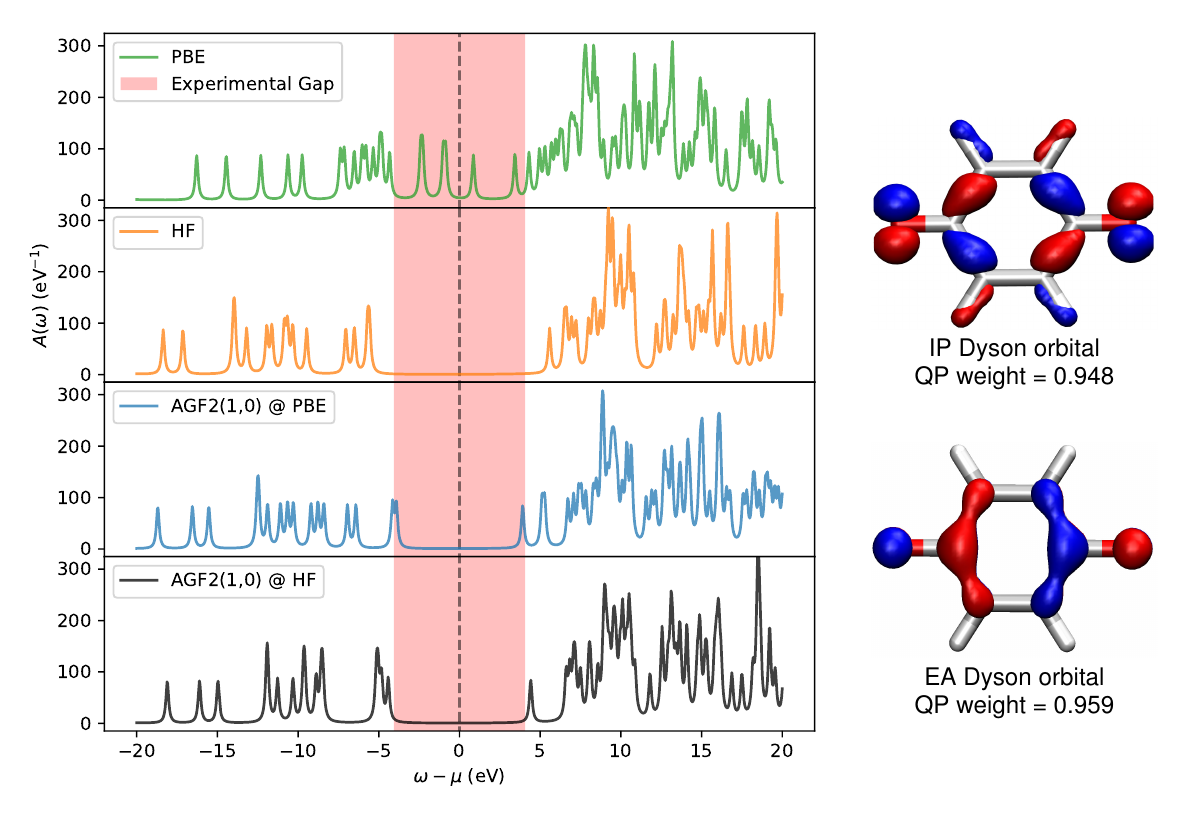}
    \caption{Quasiparticle (photoemission) spectra for the benzoquinone molecule in an aug-cc-pVDZ basis calculated at the PBE, HF and AGF2(1,0) levels \toadd{, the latter being performed using both PBE and HF as initial reference states}. 
    The red shaded area denotes the experimental spectral gap\cite{Dougherty1977,Fu2011}, centred at $\mu$, chosen to align the center of the gaps.
    Also plotted are the AGF2(1,0)\toadd{@HF} Dyson orbitals showing the spatial localization of the first IP (upper right) and EA (bottom right) for the system, along with their transition amplitudes or quasiparticle weights.
    All spectra include a broadening of $\eta = 0.1$ eV}
    \label{fig:benzoquinone}
\end{figure}

Contained in the data set were the molecules BN and C$_2$, exhibiting small but significant amounts of static correlation, which manifests in closely degenerate electronic states in the reference wavefunction. These stronger correlation effects can be identified within AGF2 from a substantial renormalization of the one-body terms in the Hamiltonian, due to a large relaxation of the density away from the initial idempotent mean-field state. This is a similar rationale to the consideration of ${\hat T}_1$ amplitude norm in coupled-cluster methods for stronger correlation effects.
As such, performance of single-reference correlated methods on these systems is often poor, and renormalization of the correlations to higher order is expected to be important.

For BN, AGF2(1,0) outperforms the ADC methods for both IPs and EAs which are substantially in error, however does not offer improvements on EOM-CC2.
The error with respect to EOM-CCSD(T)(a)* for the IP and EA are 0.91 and 0.93 eV respectively, compared to 1.79 and 1.43 eV for ADC(2), and increasing to 3.88 and 1.61 eV for ADC(3), while EOM-CC2 performs best with an error of 0.62 and 0.44 eV.
For C$_2$ however, AGF2(1,0) improves substantially on both EOM-CC2 and the ADC methods.
The error for the IP and EA are 0.308 and 0.311 eV respectively, compared to 0.733 and 1.011 eV for ADC(2), 3.318 and 0.858 eV for ADC(3) and 0.781 and 0.893 eV for EOM-CC2.
This result indicates the advantage of the self-consistency in higher-body correlation for the AGF2 in these systems with small static correlation components, in lieu of an appropriate multireference formulation \cite{Sokolov2018}.

\toadd{To ensure that conclusions as to the accuracy of AGF2(1,0) are not biased by the relatively small basis, we increase the size of the basis to aug-cc-pVTZ in Fig.~\ref{fig:w4_summary_augccpvtz}. For these calculations, AGF2 makes use of density-fitted electronic repulsion integrals to permit the larger basis.
The same W4-11 dataset was used to produce these plots, with the exception of the removal of the SO$_2$ system, which proved a large outlier in this basis. It was found that the difference between $\Delta$CCSD(T) and EOM-CCSD(T)*(a) at this basis was over 0.35eV for the EA, and therefore we do not trust this benchmark to be sufficiently accurate to lead to robust conclusions.
Nevertheless, the data shows a very similar level of accuracy in the performance of the respective methods in calculating IPs between the basis sets, with AGF2(1,0) giving a similar accuracy to ADC(3), and substantially more accurate than CC2 and ADC(2).
For the EAs, there is a much larger reduction in error for all methods, with AGF2(1,0) now performing to a similar accuracy as CC2, ADC(2) and ADC(3).
However, AGF2(1,0) shows a particularly impressive lack of bias in error across the systems, with the mean {\em signed} error being exceptionally low, and still exhibits a mean absolute error on a similar scale to CC2 and ADC(2).
With the difference in errors between the methods being much greater for IPs than EAs, combined with the much better performance of AGF2(1,0) for IPs, one can assert that energy gaps and quasiparticle excitation levels are also much better at the AGF2(1,0) level than ADC(2) or CC2 in this basis.
Data on specific systems for this basis are available in the Supplementary Information as a scatter plot of the correlation to benchmark values.
}

\subsection{Application to benzoquinone molecule}

A significant benefit of the AGF2 approach is that since the large $2h1p$ and $1h2p$ spaces are contracted through the effective dynamical truncations into only a linear dimension with system size, the effective Hamiltonian is subsequently completely diagonalized, in a step which is only $\mathcal{O}[N^3]$ scaling. 
This allows the method to extract all the charged excitations present in the method simultaneously, giving their excitation energies and transition amplitudes in order to construct the full quasiparticle spectrum and a description of the charged dynamics at all energies.
The resulting spectrum is manifestly causal, and obeys the appropriate sum rules.

To demonstrate this, and show application to a larger system, we apply AGF2(1,0) to a benzoquinone molecule in an aug-cc-pVDZ basis (220 basis functions).
\toadd{These calculations make use of density-fitted electronic repulsion integrals.}
The geometry was optimized at the level of MP2 in a cc-pVTZ basis \cite{cccbdb}. 
This system size does not represent a limit to the capabilities of the method, which we aim to demonstrate in a future study with a more efficient and parallelized implementation of the approach.
Benzoquinone is an organic acceptor molecule, with the low-lying LUMO having $\pi$ character and the HOMO a great deal of non-bonding character due to the oxygen lone pairs.
As a result, a non-empirically-tuned PBE functional has been shown to be unable to reliably predict the IP and EA due to an unbalanced description\cite{Gallandi2016}.
Fig.~\ref{fig:benzoquinone} shows the AGF2(1,0) benzoquinone spectrum broadened with a factor $\eta = 0.1$ eV \toadd{built on both an initial HF and PBE reference state}. Also included is a plot of the spatial character of the Dyson orbitals corresponding to the IP (highest negative-energy pole) and EA (lowest positive-energy pole) \toadd{of the solution corresponding to the HF reference}. 
These are found simply from the projection of the quasi-molecular orbitals into the physical degrees of freedom. 
The quasiparticle weight of these excitations can also be calculated as the squared weight of these orbitals in the physical space. 
Since the orbitals have weight on both the auxiliary and physical degrees of freedom, this allows for a non-unit transition amplitude for each excitation, which is not possible within a mean-field picture. 
This gives a weight of 0.948 (IP) and 0.959 (EA), demonstrating a reduction in their transition amplitudes due to the correlation.

In addition, Fig.~\ref{fig:benzoquinone} shows equivalent spectra for Hartree--Fock and density functional theory using a PBE functional.
The gap is predicted to be 11.162 eV by Hartree--Fock, which is \toadd{substantially larger than the 7.831-8.897 eV gap predicted by AGF2(1,0) built on PBE and HF references respectively}. This shrinking of the bandgap as a result of the truncated effective dynamics in the self-consistent AGF2(1,0) is in good agreement with experimental values of 8.14 eV \cite{Dougherty1977,Fu2011}. In contrast,
PBE predicts a dramatically smaller band gap of 1.711 eV from the Kohn-Sham hamiltonian, however with the lack of Koopmans' theorem in DFT these eigenvalues are not physically meaningful for the prediction of IPs and EAs.
As a result, many DFT functionals have been shown to fail in the quantitative agreement of orbital energies with experiment for functionals without long-range corrections \cite{McKechnie2015}.

\toadd{Due to the truncation of the dynamics in AGF2(1,0), the self-consistent solution is not entirely independent of the reference, which is shown by the slightly differing spectra when starting from HF and Kohn-Sham (PBE) reference states.
With truncations of higher orders, AGF2 will become independent of this reference, and in the limit of dynamical GF2 is (in theory) completely independent.
However, even at this low (1,0) truncation, the spectrum is certainly quantitatively similar between these different initial starting points, even for the energetics of core states, and this is certainly an appealing feature, due to the self-consistency in the method.
}


\section{Conclusions}

In applying the AGF2 approach to the computation of charged excitation spectra, we find that retaining a self-consistent formalism, while systematically truncating the resolution of the dynamics of the effective self-energy arising from the $2h1p$ and $1h2p$ spaces can result in a highly accurate and robust approach. 
Of these truncations, AGF2(1,0) has been shown to outperform many other similarly-scaling quantum chemical approaches for both IPs and EAs in a $\mathcal{O}[N^5]$ scaling method. 
It was shown across the W4-11 test set to have an accuracy close to that of the higher scaling ADC(3) method, in a comparison against EOM-CCSD(T)(a)*. 
We also demonstrated the ability to compute entire spectra in a single-shot, rather than requiring a state-specific optimization, and computed the correlated quasiparticle spectrum of the benzoquinone molecule. 
We expect this approach to offer a very competitive method, and future work will implement an efficient and parallel algorithm, in order to extend the scope of the approach to open problems in theoretical photoemission spectroscopy, while similar ideas are also being explored for the application to neutral excitation spectra.


\section{Acknowledgements}

The authors sincerely thank Alejandro Santana-Bonilla for technical help, as well as Max Nusspickel for useful discussions.
G.H.B. also gratefully acknowledges support from the Royal Society via a University Research Fellowship, as well as funding from the European Union's Horizon 2020 research and innovation programme under grant agreement No. 759063. We are grateful to the UK Materials and Molecular Modelling Hub for computational resources, which is partially funded by EPSRC (EP/P020194/1).

\ifjcp
%
\else
\providecommand{\latin}[1]{#1}
\makeatletter
\providecommand{\doi}
  {\begingroup\let\do\@makeother\dospecials
  \catcode`\{=1 \catcode`\}=2 \doi@aux}
\providecommand{\doi@aux}[1]{\endgroup\texttt{#1}}
\makeatother
\providecommand*\mcitethebibliography{\thebibliography}
\csname @ifundefined\endcsname{endmcitethebibliography}
  {\let\endmcitethebibliography\endthebibliography}{}

\fi

\end{document}